\documentclass[runningheads]{llncs}
\usepackage{graphicx}
\usepackage{booktabs}
\usepackage{xcolor}
\usepackage{graphicx}
\usepackage{verbatim}
\usepackage{comment}
\usepackage[colorinlistoftodos,prependcaption,textsize=tiny]{todonotes}

\begin{document}
\newcommand\PDT[1]{\todo[linecolor=blue,backgroundcolor=blue!25,bordercolor=blue]{#1}}
\newcommand\CSF[1]{\todo[linecolor=red,backgroundcolor=red!25,bordercolor=red]{#1}}
\newcommand\HMO[1]{\todo[linecolor=green,backgroundcolor=green!25,bordercolor=green]{#1}}
\newcommand\MSG[1]{\todo[linecolor=pink,backgroundcolor=pink!25,bordercolor=pink]{#1}}
\title{Test-time Unsupervised Domain Adaptation}
%
%
\author{Thomas Varsavsky\inst{1}${}^{,}$\inst{2}
\and Mauricio Orbes-Arteaga\inst{1}${}^{,}$\inst{3}
\and Carole H. Sudre\inst{1}${}^{,}$\inst{4}
\and Mark S. Graham\inst{1}
\and Parashkev Nachev\inst{5}
\and M. Jorge Cardoso\inst{1}} 

\institute{
 School of Biomedical Engineering and Imaging Sciences, KCL, UK
\and Department of Medical Physics and Biomedical Engineering, UCL, UK\\
\and Biomediq A/S, Copenhagen, Denmark \\
\and Dementia Research Centre, Institute of Neurology, UCL, London, UK \\
\and High Dimensional Neurology Group, Institute of Neurology, UCL, London, UK}%
\authorrunning{Thomas Varsavsky}
%

\maketitle              
\begin{abstract}
 Convolutional neural networks trained on publicly available medical imaging datasets (source domain) rarely generalise to different scanners or acquisition protocols (target domain). This motivates the active field of domain adaptation. While some approaches to the problem require labelled data from the target domain, others adopt an unsupervised approach to domain adaptation (UDA). Evaluating UDA methods consists of measuring the model's ability to generalise to unseen data in the target domain. In this work, we argue that this is not as useful as adapting to the test set directly. We therefore propose an evaluation framework where we perform test-time UDA on each subject separately. We show that models adapted to a specific target subject from the target domain outperform a domain adaptation method which has seen more data of the target domain but not this specific target subject. This result supports the thesis that unsupervised domain adaptation should be used at test-time, even if only using a single target-domain subject.

\end{abstract}
\section{Introduction}
Recent years have seen huge progress in performance in brain MRI segmentation, classification and synthesis largely thanks to the application of convolutional neural networks to these problems. The organisation of challenges such as BRATS \cite{menze2014multimodal} and the MICCAI 2017 White Matter Hyperintensity Challenge \cite{kuijf2019standardized} have allowed the community to benchmark their segmentation algorithms on research data. In these cases, training data is usually  preprocessed following a consistent protocol with techniques such as skull stripping, bias field correction, histogram normalisation and co-registration. Efforts are often put in place to ensure a certain degree of standardisation across the centres providing data, in terms of scanners parameters such as field strength, manufacturer, echo time, relaxation time and contrast agent. In addition, individuals generally have similar pre-clinical conditions and pathological presentations. When applied to data from clinical practice that presents much more heterogeneous acquisition conditions, the performance of algorithms trained on challenge data degrades. Performance can improve if algorithms are fine-tuned on labelled data in the target domain, but these can be expensive to acquire and rely on relative homogeneity of acquisition parameters in the target domain. If no labels are available then unsupervised domain adaptation may be used, which has seen growing interest in recent years e.g. \cite{kamnitsas2017unsupervised,perone2019unsupervised}. 

Domain is not always a clear binary label. Scans of a particular MR modality (e.g T1-weighted) may come from the same scanner in the same hospital but may use different acquisition parameters. Variability can be so large that each image can almost be considered its own domain.

When evaluating domain adaptation methods for segmentation, there is often a training set, a validation set and a test set for both source and target domains. Methods are judged on their ability to generalise from seen data in the source domain to unseen data in the target domain. In this work we argue for a different evaluation criterion, namely how well a model performs on the data it adapts to. We call this \textit{``test-time unsupervised domain adaptation''}. When this test-time adaptation is performed on an individual subject we call it \textit{``one-shot unsupervised domain adaptation''}. We present a domain adaptation method which leverages a combination of adversarial learning and consistency under augmentation to work in this one-shot case. We apply this methodology on multiple sclerosis lesion segmentation but it is designed to be applicable to other tasks in medical imaging.


\subsubsection{Related work:}
Our work considers the use of existing unsupervised domain adaptation methods when only a single \textit{unlabelled} sample from the target domain is available. In this work we use the same data, pre-processing and segmentation task as in \cite{valverde2019one}, where the authors tackle one-shot supervised domain adaptation, adapting to a target domain using a single \textit{labelled} subject.

It is worth mentioning the framework proposed by Zhao et al \cite{zhao2019data} and highlighting the difference to this work. The authors consider the variability between single-modality brain MRIs to be quantifiable by an additive intensity transform and a spatial transform to a brain atlas. They use this technique to create an entire labelled dataset from a single brain with an associated anatomical parcellation (hence the term ``one-shot''). While the intensity transform tackles the variation in acquisition parameters, the spatial transform covers variations in anatomy. Although this and follow-up work produce realistic training data in the context of brain parcellation, such scheme cannot be trivially extended to application to pathologies in which the variability in presentation, location and extent is far greater. This is especially true in lesion segmentation, where a lesion prior cannot be produced from a non-linear deformations of an atlas. \looseness=-1

Neural style-transfer methods were recently applied for unsupervised domain adaptation of cardiac MRI in \cite{ma2019neural}. The style of the target domain is matched to that of a single subject in the source domain by simultaneously minimising a style loss $l_{style}(\hat{y}, y)$ and a content loss $l_{content}(\hat{y}, x)$ where $\hat{y}$ is the generated style-transferred image, $x$ is the image from the target domain and $y$ is the image from the source domain. This method relies on finding an image in the source domain which most closely resembles the target image based on a Wasserstein distance metric. This method is similar to ours in that adaptation is performed on each individual test subject as its own optimisation problem.

Recent advances in self-supervised learning have led to large improvements in semi-supervised learning. Methods such as \cite{carlucci2019domain} use self-supervised tasks such as solving jigsaw puzzles to perform domain adaptation. Promoting invariance in networks outputs under data augmentation is another self-supervised task which was shown to work well for domain adaptation in \cite{french2017self} and which we refer to as Mean Teacher. It was adapted for use in medical image segmentation in \cite{perone2019unsupervised}. In \cite{orbes_and_varsavsky} the authors showed improvements over Mean Teacher using a simpler paired consistency method. They used paired data as a form of ``ground-truth augmentation''. When paired data is not available, which is most common in practice, small adjustments to this method can lead to substantial improvements. The method of \cite{orbes_and_varsavsky} was chosen to demonstrate the value of test-time UDA, as it reported better results than domain adversarial learning and Mean Teacher on a related task. However, note that our domain adaptation methodology is not bound to a particular method.

\section{Domain Adversarial Learning and Paired Consistency}
We adapt the method for domain adaptation described in \cite{orbes_and_varsavsky} which consists of domain adversarial learning and consistency training. In domain adversarial training we seek to find a feature representation $\phi_{\theta}(x)$ which contains as little information about $d$ - the domain of $x$ - as possible and the most information about the label $y$. We do so by including a domain discriminator $D_{\gamma}(x)$ which predicts a domain $\hat{d}$ and is trained by minimising the binary cross-entropy between this prediction and the ground-truth domain $d$, $\mathcal{L}_{adv} = l_{bce}(D_{\gamma}(\phi_{\theta}(x)), d)$. We use the gradient reversal layer from \cite{ganin2016domain} to guarantee that the network weights $\theta$ change in the direction which minimises the supervised loss $\mathcal{L}_{sup}$ and maximises the adversarial loss $\mathcal{L}_{adv}$ where $\mathcal{L}_{sup} = l(\mathcal{M}(x), y_s)$ (we use the dice loss for $l$). 

Consistency training is a simple semi-supervised learning method which works by enforcing invariance to data augmentation. A model $\mathcal{M}$ is trained to produce a prediction $\hat{y}_s$ on some source data $x_s$ which has an associated label $y_s$ using a regular supervised loss $\mathcal{L}_{sup}$. An image from the target domain $x_T$ is passed to the same model $\mathcal{M}$ to obtain $\hat{y}_T$. The same image is passed through the model after augmentation $g(x_T)$ (details about the choice of $g$ in section \ref{section:aug}) to produce $\hat{y}^{aug}_{T}$. The paired consistency loss $\mathcal{L}_{pc}$ aims at minimising the difference between $\hat{y}_T$ and $\hat{y}^{aug}_{T}$. Following the guidance from \cite{perone2019unsupervised} and \cite{orbes_and_varsavsky}, the soft dice is used as $\mathcal{L}_{pc}$, defined as $\mathcal{L}_{pc}(\hat{y},\hat{y}^{aug}) = \sum_{i=1}^{N} \hat{y}_i\hat{y}_i^{aug} / (\sum_{i=1}^{N} \hat{y}_i + \sum_{i=1}^{N} \hat{y}_i^{aug})$. By enforcing predictions to be invariant to some noise of perturbation $\delta$, $y(x) = y(x+\delta)$, we encourage the decision boundary of our classifier to fall in regions of low density

\begin{figure}[h]
    \centering
    \includegraphics[scale=0.05]{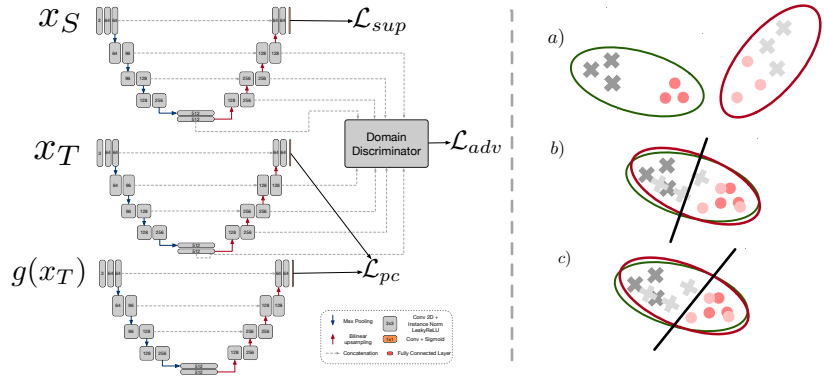}
    \caption{Left: Our domain adaptation method uses a paired consistency loss $\mathcal{L}_{pc}$ which encourages predictions from the target image $x_T$ to be invariant to some augmentation $g$. The backbone is a single 2D U-Net (parameters are shared) with features from each depth being interpolated bilinearly, concatenated and fed to a domain discriminator which uses an adversarial loss $\mathcal{L}_{adv}$ to maximise domain confusion. Right: In a) we depict representations of pixels in some feature space, the green circle is source and red target with crosses and circles depicting foreground and background. b) shows what happens when we introduce an adversarial loss, the feature spaces are shifted such that they are indistinguishable from the source domain but the decision boundary is drawn with only souce data. In c) we show the effect of the PC loss in moving the decision boundary to an area of low-density}
    \label{fig:method}
\end{figure}

The right hand side of Figure \ref{fig:method} (right) depicts the benefits of domain adversarial learning. In frame a) we see a source and target domain represented by green and red ovals respectively. They contain representations of foreground and background pixels shown as grey crosses and red dots. Frame b) shows what happens when domain adversarial learning is used. The domains become indistinguishable which makes the ovals overlap. However, when the decision boundary is drawn to separate the two classes it is done only by looking at the source domain. In frame c) we introduce paired consistency. The unlabelled points are near the labelled ones, they will be assigned the label of their nearest cluster which allows the boundary to be redrawn in an area of low density. We include some t-SNE plots of our learned features in Figure 3 of the Supplementary Material which clearly show the positive effect of domain adaptation to the separability of lesion and background across both domains.

The method proposed in \cite{orbes_and_varsavsky} achieved consistency training using what they denote as ``ground-truth augmentation''. This means two registered scans of the same patient using different acquisition parameters. In this work, we avoid this requirement by providing stronger augmentation and dropping the third output of their domain discriminator which sought to find a feature space which contained no information about whether an image was source, target or target augmented. Note that this minor change significantly reduces the data requirements of the model.

\subsubsection{Implementation details}
We use a simple 2D U-Net with five levels as the backbone of our model. Each encoding block has two 2D convolutions with kernel sizes of $3\times3$, a stride of 1, and padding of 1 (except the first which has a padding of 2 and kernel size 5). The blocks have gradually increasing number of filters: 64, 96, 128, 256, 512. We use instance norm and leaky ReLU after each convolution in each block as in \cite{isensee2018nnu}. We use max pooling between each encoder block and bilinear upsampling between each decoder block and the standard concatenation of feature vectors from the same depth.

For the domain discriminator we use a small VGG-style convolutional neural network with four convolutions of kernel size 3 and stride of 2 each followed by a batch norm operation and three fully connected layers of size 28800, 256 and 128 respectively with 0.5 dropout in between. We follow the suggestion from \cite{kamnitsas2017unsupervised} to feed in a concatenated vector of multi-depth features as input to the discriminator. Specifically, we take the activations from each depth of the decoder (excluding the center of the U-Net) and use bilinear interpolation to make them the same shape as the penultimate depth in the spatial dimension. We then concatenate on the channels dimension. All code is written in PyTorch and will be made available at the time of publication.

\section{Experiments}
In the proposed test-time UDA, an unusual approach to train\/val\/test splits is taken. In fact, part of the data for which we train the paired consistency component of our model $\mathcal{M}$ is the one on which the labelling quality is tested. Please note that the labels of the test set are \textit{never} used during training. In order to prevent data leakage, all hyperparameters tuning strategies and model selection steps were performed on a completely separate dataset (results not shown). 
Each UDA run was trained for exactly 15,000 iterations using a batch size of 20 with the exception of the supervised baseline which had a validation subject to allow for model selection. We used the Adam optimiser with a learning rate of $1\times10^{-3}$ with no learning rate policy. A separate Adam optimiser with learning rate $1\times10^{-4}$ was used for the discriminator. In order to further validate our model we submit results to the online validation server for the ISBI 2015 challenge. We provide results for the first timepoint of each of the test subjects in the supplementary material.

\subsubsection{Augmentation}
\label{section:aug}

 In \cite{perone2019unsupervised} the authors used random affine transforms (rotating, scaling, shearing and translating) as well as random elastic deformations where an affine grid is warped and applied to the image. Their method  does augmentation on the output of a neural network but this output does not need to be differentiated. We use all these augmentations but exclude elastic deformation, as it is difficult to implement in a differentiable manner (a requirement of the proposed method). Following the recommendations in \cite{orbes_and_varsavsky} we use augmentation which is realistic, valid and smooth. To this end, we also add bias field augmentation \cite{gibson2017niftynet} and k-space augmentation \cite{shaw2019mri} as extra transformations, as they have been shown to produce realistic variations in MRIs.

\subsubsection{Data}

Domain adaptation is here applied to multiple sclerosis lesion segmentation as an exemplary task. We use as source domain data from two separate MICCAI challenges on multiple sclerosis lesion segmentation MS2008 \cite{styner20083d} and MS2016 \cite{commowick2018objective}. Data from ISBI2015 \cite{carass2017longitudinal} is used as target domain. The FLAIR sequences from each of these datasets are skull-stripped (using HD-BET \cite{isensee2019automated}), bias-field corrected using the N4 algorithm and registered to MNI space as in \cite{valverde2019one}.

\subsection{Results}
We present results from five different methods. First, there is a lower bound provided by using a model trained on the source domain and applied to data from the target domain, which we refer to as \textit{no adaptation}. The highest expected performance is provided by training a model on the target domain images and labels, fine-tuned from a model trained on the source domain, which we refer to as \textit{supervised}. When we use paired consistency and adversarial learning to domain adapt to a single subject on the target domain, this is denoted as \textit{One-shot UDA}. We compare this against a model which sees this and two more subjects from the target domain, and refer to it as \textit{Test-time UDA}. A comparison was also made against a traditional approach to domain adaptation where the model trains on target domain data which excludes the test subject; we refer to this variant as \textit{Classic UDA}. In Table 1 of the supplementary material we show results for each of these methods evaluated on a variety of metrics. These were chosen to match those in \cite{carass2017longitudinal}. The LFPR is the lesion false positive rate and LTPR is the lesion true positive rate which are implemented as in \cite{styner20083d}. We follow the recommendations of the MICCAI Grand Challenges, specifically the method described in \cite{simpson2019large}, to provide a single rank score comparing all methods. Note that this ranking method provides a single summary metric that incorporates a per-metric non-parametric statistical significance model.

\begin{table}
\centering
\scalebox{.77}{
\begin{tabular}{l|c|c|c|c|c|c|c|c}
\toprule
        Method &  Rank &            Dice &         Hausdorff &             LFPR &            LTPR &             PPV &      Sensitivity &        Vol Diff \\
\midrule
    Supervised &  1.71 &  0.67 $\pm$ 0.1 &      37. $\pm$ 8. &   0.52 $\pm$ 0.2 &  0.61 $\pm$ 0.2 &  0.67 $\pm$ 0.2 &   0.73 $\pm$ 0.2 &  0.44 $\pm$ 0.2 \\
    Test-time UDA (ours) &  2.43 &  0.61 $\pm$ 0.2 &      48. $\pm$ 5. &   0.54 $\pm$ 0.2 &  0.57 $\pm$ 0.2 &  0.54 $\pm$ 0.2 &  0.76 $\pm$ 0.09 &   0.72 $\pm$ 1.0 \\
    One-shot UDA (ours) &  2.71 &  0.60 $\pm$ 0.2 &  47. $\pm$ 11 &   0.52 $\pm$ 0.2 &  0.51 $\pm$ 0.2 &  0.54 $\pm$ 0.2 &  0.76 $\pm$ 0.09 &   0.92 $\pm$ 2.0 \\
    Classic UDA &  3.86 &  0.56 $\pm$ 0.1 &  47. $\pm$ 14 &   0.55 $\pm$ 0.2 &  0.58 $\pm$ 0.2 &  0.49 $\pm$ 0.2 &   0.73 $\pm$ 0.2 &  0.78 $\pm$ 0.5 \\
 No adaptation &  4.29 &  0.57 $\pm$ 0.1 &      55. $\pm$ 7. &  0.68 $\pm$ 0.08 &  0.55 $\pm$ 0.2 &  0.49 $\pm$ 0.2 &  0.76 $\pm$ 0.08 &  0.76 $\pm$ 0.7 \\
\bottomrule
\end{tabular}
}

\caption{Results on metrics described in \cite{carass2017longitudinal}. The metrics are ranked using the scheme from \cite{simpson2019large} to provide a rank score. The proposed test-time methods are labelled (ours).}

\label{tab:ranking_results}
\end{table}

\begin{figure}[ht]
    \centering
    \includegraphics[width=\textwidth]{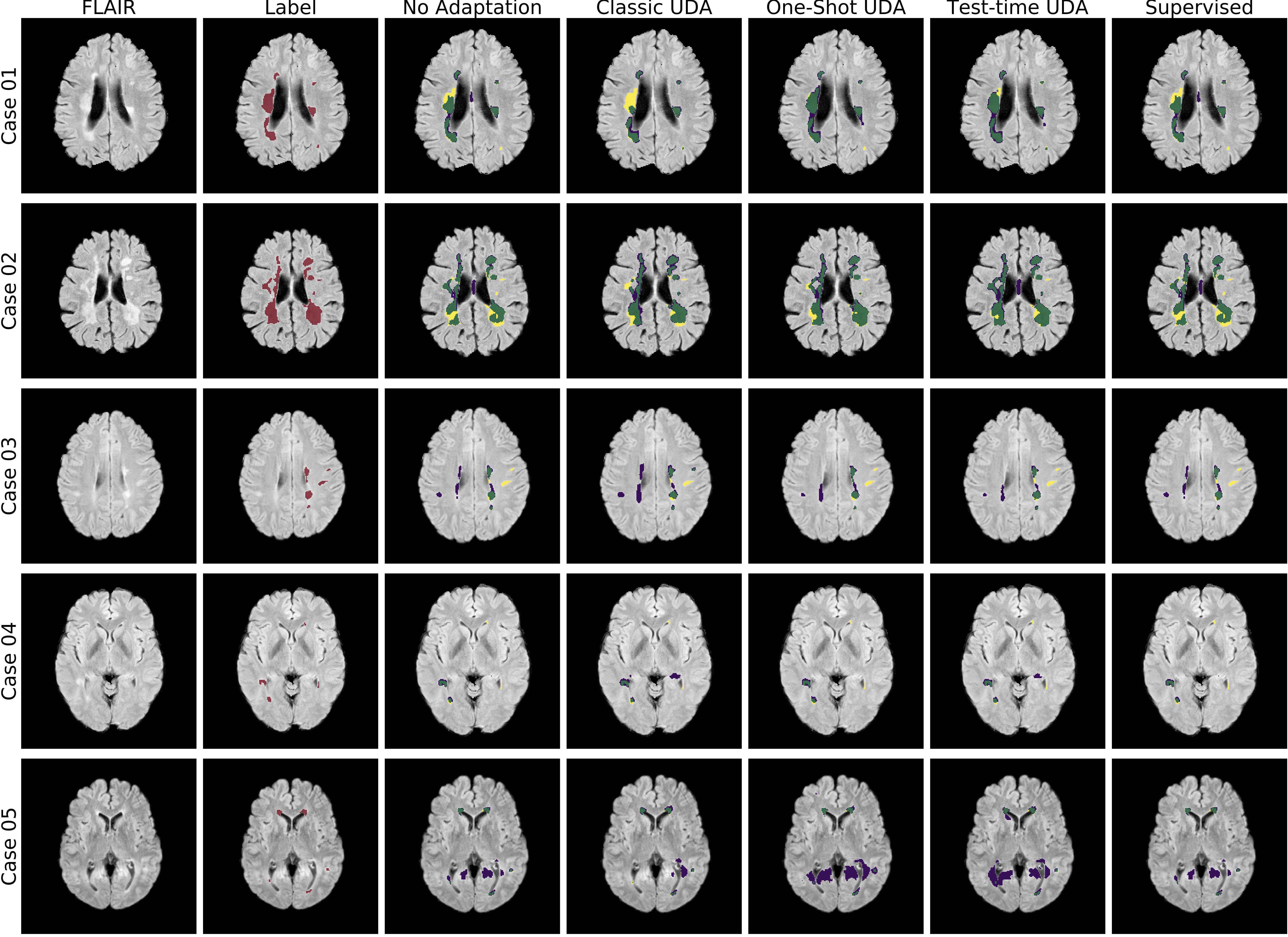}
    \caption{Some qualitative results comparing no adaptation, classic unsupervised domain adaptation, one-shot unsupervised domain adaptation, test-time unsupervised domain adaptation, and the hypothetical gold-standard using supervised learning. Red denotes the ground-truth annotation, true positives are shown in green, false negatives are in yellow and false positives are in blue.}
    \label{fig:my_label}
\end{figure}

\section{Discussion}

The results in Table \ref{tab:ranking_results} show a clear ordering with Supervised as the best performing method, as expected, followed by test-time UDA, one-shot UDA, classic UDA and finally no adaptation. These results reveal that learning enough information about a domain shift, i.e. Classic UDA, is not enough to get the best performance on each test subject in the target domain. By domain-adapting to each test subject, we are adapting to the subjects individual anatomical and pathological presentation. It is also worth mentioning that our One-shot unsupervised domain adaptation achieved a dice of 0.60 on the ISBI training set which is comparable to the 0.58 reported  on the ISBI holdout set in \cite{valverde2019one} despite not using a single label from ISBI. Results in Table \ref{tab:my_label} show the performance of Test-time UDA against a Supervised baseline, Classic UDA and One-shot UDA. Classic UDA outperformed One-shot, but test-time UDA was best of all. Future work will include experiments on brain tumour segmentation and compare additional UDA methods in the Classic, One-shot and Test-time settings.
\begin{table}[h]
    \centering
    \scalebox{.77}{\begin{tabular}{l|c|c|c|c|c|c|c}
\toprule
           Method &  Rank &            Dice &            LFPR &            LTPR &             PPV &             TPR &              Volume Difference \\
\midrule
         Valverde et al. (Supervised) &  1.50 &  0.60 $\pm$ 0.2 &  0.22 $\pm$ 0.2 &  0.41 $\pm$ 0.2 &  0.73 $\pm$ 0.2 &  0.54 $\pm$ 0.2 &  5829 $\pm$ 7900 \\
            Test-time UDA (ours) &  4.25 &  0.51 $\pm$ 0.2 &  0.53 $\pm$ 0.2 &  0.25 $\pm$ 0.2 &  0.59 $\pm$ 0.2 &  0.51 $\pm$ 0.2 &  6947 $\pm$ 8800 \\
             Classic UDA &  4.42 &  0.49 $\pm$ 0.2 &  0.54 $\pm$ 0.2 &  0.28 $\pm$ 0.2 &  0.55 $\pm$ 0.2 &  0.48 $\pm$ 0.2 & 5784 $\pm$ 7500 \\
          One-shot UDA (ours) &  4.50 &  0.48 $\pm$ 0.2 &  0.52 $\pm$ 0.3 &  0.28 $\pm$ 0.1 &  0.52 $\pm$ 0.3 &  0.51 $\pm$ 0.2 &  7009 $\pm$ 7700 \\
\bottomrule
\end{tabular}}
    \caption{Results from the ISBI 2015 holdout set hosted at \url{https://smart-stats-tools.org/lesion-challenge}. We ran our three UDA methods on the first timepoint of each of the 14 test subjects. Note that one of the limitations of this form of validation is the low inter-rater disagreement reported in Carass et al. The same ranking scheme was used as in the training set, however the symmetric distance was used instead of the Hausdorff. The Classic UDA outperformed One-shot but test-time UDA was best of all.}
    \label{tab:my_label}
\end{table}


\section{Conclusion}
Existing approaches to unsupervised domain adaptation in medical image segmentation adapt to subjects in a target domain. The performance of these algorithms is then measured based on how well they generalise to unseen subjects in this target domain. When looking through scans in a hospital PACS system there is a large amount of heterogeneity in acquisition parameters. As an example, at our local hospital (anonymous), we found more than 1400 different brain MRI sequences being used. We can thus think of each of these scans as its own domain, which motivates what we call ``test-time unsupervised domain adaptation''. Note that this is not an algorithmic modification, but simply a training and testing framework, where a domain adaptation algorithm is trained and evaluated on the same target data. We perform experiments using a modern domain adaptation technique which combines the benefits of domain adversarial learning and consistency regularisation. Our experiments on multiple sclerosis lesions suggest that using domain adaptation on a single subject can be more effective than classic domain adaptation on more subjects. 
%

%
%
\bibliographystyle{splncs04}
\bibliography{biblio}
\end{document}